\documentclass[twocolumn,superscriptaddress]{revtex4}
\usepackage{graphicx}
\usepackage{dcolumn}
\usepackage{amsmath}
\usepackage{amsfonts}
\usepackage{amssymb}

\begin{document}

\title{\textbf{A quantitative model of trading and price formation in
financial markets}}

\author{Marcus G. Daniels}
\affiliation{Santa Fe Institute, 1399 Hyde Park Rd., Santa Fe NM 87501}
\email{mgd@santafe.edu}

\author{J. Doyne Farmer}
\affiliation{Santa Fe Institute, 1399 Hyde Park Rd., Santa Fe NM 87501}
\email{jdf@santafe.edu}

\author{L\'{a}szl\'{o} Gillemot}
\affiliation{Santa Fe Institute, 1399 Hyde Park Rd., Santa Fe NM 87501}
\email{laci@santafe.edu}

\author{Giulia Iori}
\affiliation{Mathematics Dept. Kings College, London, Strand, London WC2R 2LS}
\email{giulia.iori@kcl.ac.uk}

\author{Eric Smith}
\affiliation{Santa Fe Institute, 1399 Hyde Park Rd., Santa Fe NM 87501}
\email{desmith@santafe.edu}

\date{January 29, 2002}

\begin{abstract}
We use standard physics techniques to model trading and price
formation in a market under the assumption that order arrival and 
cancellations are Poisson random processes.  This model makes testable
predictions for the most basic properties of a market, such as the
diffusion rate of prices, which is the standard measure of financial risk,
and the spread and price impact functions, which are the main determinants
of transaction cost.  Guided by dimensional analysis, simulation, and
mean field theory, we find scaling relations in terms of order flow
rates.  We show that even under completely random order flow the need
to store supply and demand to facilitate trading induces anomalous
diffusion and temporal structure in prices.
\end{abstract}
\maketitle

There have recently been efforts to apply physics methods to problems
in economics \cite{Mantegna00,Bouchaud00}.  This effort has yielded
interesting empirical analyses and conceptual models.  However, with
the exception of refinements to option pricing theory, so far it has
had little success in producing theories that make falsifiable
predictions about markets.  In this paper we develop a mechanistic
model of a standard method for trade matching in order to make
quantitative predictions about the most basic properties of markets.
This model differs from standard models in economics in that we make
no assumptions about agent rationality.  The model makes falsifiable
predictions based on parameters that can all be measured in real data,
and preliminary results indicate that it has substantial explanatory
power \cite{Daniels02}

The random walk model was originally introduced by Bachelier to
describe prices, five years before Einstein used it to model Brownian
motion \cite{Bachelier00}. Although this is one of the most widely
used models of prices in financial economics, there is still no
understanding of its most basic property, namely, its diffusion rate.
We present a theory for how the diffusion rate of prices depends on
the flow of orders into the market, deriving scaling relations based
on dimensional analysis, mean field theory, and simulation.  We also
make predictions of other basic market properties, such as the gap
between the best prices for buying and selling, the density of stored
demand vs. price, and the impact of trading on prices.

Most modern financial markets operate continuously.  The mismatch
between buyers and sellers that typically exists at any given instant
is solved via an order-based market with two basic kinds of
orders. Impatient traders submit
\textit{market orders}, which are requests to buy or sell a given
number of shares immediately at the best available price. More patient
traders submit \textit{limit orders}, which also state a limit price,
corresponding to the worst allowable price for the transaction.  Limit
orders often fail to result in an immediate transaction, and are
stored in a queue called the \textit{limit order book}. Buy limit
orders are called \textit{bids}, and sell limit orders are called
\textit{offers }or \textit{asks}. We will label the logarithm of the best 
(lowest) offering price $a(t)$ and the best (highest) bid price $b(t)$. 
There is typically a non-zero price gap between them, called the
\textit{spread} $s(t)=a(t)-b(t)$.  

As market orders arrive they are matched against limit orders of the
opposite sign in order of price and arrival time. Because orders are
placed for varying numbers of shares, matching is not necessarily
one-to-one. For example, suppose the best offer is for 200 shares at
\$60 and the the next best is for 300 shares at \$60.25; a buy market 
order for 250 shares buys 200 shares at \$60 and 50 shares at
\$60.25, moving the best offer $a(t)$ from \$60 to \$60.25.  A high
density per price of limit orders results in high
\textit{liquidity} for market orders, i.e., it implies a small price
movement when a market order of a given size is placed.

\begin{figure}[ptb]
   \begin{center} 
   \includegraphics[scale=0.45]{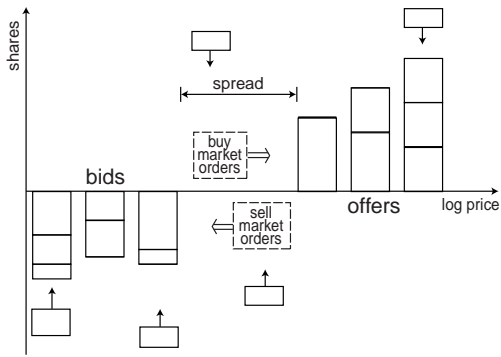}
   \caption{Schematic of the order-placement process.  Stored limit orders are
shown ``stacked'' along the price axis, with bids (buy limit orders)
negative and offers (sell limit orders) positive.  New limit orders
are visualized as ``falling'' randomly onto the price axis.  New
offers can be placed at any price greater than the best bid, and new
bids can be placed at any price less than the best offer.  Limit orders
can be removed by random spontaneous deletion or by market
orders of the opposite sign.}
   \label{orderprocess} 
   \end{center}
\end{figure}

We analyze the queueing properties of such order-matching algorithms
with the simple random order placement model shown in
Fig.~\ref{orderprocess}.
All the order flows are modeled as Poisson processes.  We assume that
market orders in chunks of $\sigma$ shares arrive at a rate of $\mu$
shares per unit time, with an equal probability for buy
orders and sell orders.  Similarly, limit orders in
chunks of $\sigma$ shares arrive at a rate of $\alpha$ shares per unit
price and per unit time.  Offers are placed with uniform probability
at integer multiples of a tick size $dp$ in the range $b(t)<p<\infty$,
and similarly for bids on $-\infty<p<a(t)$.  $p$ represents the
logarithm of the price, and $dp$ is a logarithmic price interval
\footnote{In real markets the price interval $dp$ is linear rather
than logarithmic; as long as $p \gg dp$ this is a good assumption.}. 
(To avoid repetition the word
{\it price} will henceforth refer to the logarithm of price.)  When a
market order arrives it causes a transaction.  Under the assumption of
constant order size, a buy market order removes an offer at price
$a(t)$, and a sell market order removes a bid at price
$b(t)$. Alternatively, limit orders can be removed spontaneously by
being canceled or by expiring. We model this by letting any order be
removed randomly with constant probability $\delta$ per unit time.

This order placement process is designed to permit an analytic
solution. The model builds on previous work modeling the continuous
double auction
\cite{Domowitz94,Bollerslev94,Eliezer98,Maslov00,Iori01,Challet01,
Slanina01}.  While the assumption of limit order placement over an
infinite interval is clearly unrealistic \cite{Bouchaud02,Zovko02}, it
provides a tractable boundary condition for modeling the behavior of
the limit order book in the region of interest, near the midpoint
price $m(t)=(a(t)+b(t))/2$. It is also justified because limit orders
placed far from the midpoint usually expire or are canceled before
they are executed.  For our analytic model we use a constant order
size $\sigma$.  In simulations we also use variable order size,
e.g. half-normal distributions with standard deviation
$\sqrt{2/\pi}\sigma$, which gives similar results.

For simplicity in our model we do not directly allow limit orders that
cross the best price.  For example, a buy order of size $x + y$ may
have a limit price that is higher than the best ask, so that $x$
shares immediately result in a trade, and $y$ shares remain on the
book.  Such an order is indistinguishable from a market order for $x$
shares immediately followed by a non-crossing limit order for $y$
shares. By definition our model implicitly allows such events, though
it neglects the resulting correlation in order placement.

Dimensional analysis simplifies this problem and provides rough
estimates of its scaling properties.  The three fundamental dimensions
are {\it shares}, {\it price}, and {\it time}.  There are five
parameters: three order flow rates and two discreteness parameters.
The {\it order flow rates} are $\mu$, the market order arrival rate,
with dimensions of {\em shares per time}; $\alpha$, the limit order
arrival rate per unit price, with dimensions of {\em shares per price
per time}; and $\delta$, the rate of limit order decays, with
dimensions of {\em 1/time}.  The two {\it discreteness parameters} are
the price tick size $dp$, with dimensions of {\em price}, and the
order size $\sigma$, with dimensions of {\em shares}.  Because there
are five parameters and three dimensions, and because the
dimensionality of the parameters is sufficiently rich, all the
properties of the limit-order book can be described by functions of
two independent parameters.

We perform the dimensional reduction by taking advantage of the fact
that the effect of the order flow rates is primary to that of the
discreteness parameters.  This leads us to construct nondimensional
units based on the order flow parameters alone, and take
nondimensionalized versions of the discreteness parameters as the
independent parameters whose effects remain to be understood.  There
are three order flow rates and three fundamental dimensions.
Temporarily ignoring the discreteness parameters, there are unique
combinations of the order flow rates with units of shares, price, and
time.  These define a characteristic number of shares $N_c = \mu / 2
\delta$, a characteristic price interval $p_c = {\mu} / 2 {\alpha}$,
and a characteristic timescale $t_c = 1/\delta$. (The factors of two
are a matter of convenience; they occur because we have defined the
market order rate for either a buy or a sell order to be $\mu/2$.)
These characteristic values can be used to define nondimensional
coordinates $\hat{p} = p/p_c$ for price, $\hat{N} = N/N_c$ for shares,
and $\hat{t} = t/t_c$ for time.

A nondimensional scale parameter based on order size is constructed by
dividing the typical order size $\sigma$ (which is measured in shares)
by the characteristic number of shares $N_c$, i.e. $\epsilon \equiv
\sigma/N_c = 2 \delta \sigma / {\mu}$.  $\epsilon$ characterizes the 
granularity of the orders stored in the limit order book. A
nondimensional scale parameter based on tick size is constructed by
dividing by the characteristic price, i.e. $dp/p_c = 2 {\alpha} dp /
{\mu} $.  The theoretical analysis and the simulations show that there
is a sensible continuum limit as the tick size $dp \rightarrow 0$, in
the sense that there is non-zero price diffusion and a finite spread.
Furthermore, the dependence on tick size is usually weak, and for many
purposes the limit $dp \rightarrow 0$ approximates the case of finite
tick size fairly well.

Space constraints do not permit us to review the theoretical
development of the model in this Letter; it is presented in detail in
Ref.~\cite{Smith02}.
We write an approximate master equation for the number of
shares at each price level $p$ at time $t$, and then find a
self-consistent mean field theory steady-state solution.  We also
develop an independent interval approximation, borrowing methods from
the study of reaction-diffusion equations \cite{Majumdar00}.  We find that
the theory fits the simulation results accurately for large values of
$\epsilon$. For small values of $\epsilon$ the theory continues to
capture the mean spread very well.  The predictions of other properties
remain qualitatively correct, but are no longer quantitatively
accurate.  The results we quote here are all from simulations; to see
the development of the theory and comparisons to simulation, see
ref.~\cite{Smith02}.

In the following we explore the predictions of the model for the basic
properties of markets.  As already noted, neglecting the effects of
the discreteness parameters, gives three dimensional quantities and
three parameters, which we call the {\it continuum approximation}.
For the continuum approximation dimensional analysis alone yields
simple estimates for the most relevant economic properties of the
models.  We have refined these estimates by simulation and mean
field theory, which take the effects of $\epsilon$ and $dp/p_c$ into
account.  The results are summarized in Table~\ref{scalingTable},
and described in more detail below.
\begin{table*}
\begin{tabular}[c]{llll}% 
  {\bf Quantity} & 
  {\bf Dimensions} & 
  {\bf Continuum scaling relation} & 
  {\bf Scaling from simulation and theory} 
\\
  Asymptotic depth & $shares/price$ & 
  $d \sim \alpha/\delta$ & 
  $d = \alpha/\delta$ 
\\ 
  Spread & 
  $price$ & 
  $s \sim \mu/\alpha$ & 
  $s = (\mu/\alpha) f(\epsilon, dp/p_c)$ 
\\ 
  Slope of depth profile & 
  $shares/price^{2}$ & 
  $\lambda \sim \alpha^{2}/\mu\delta = d/s$ & 
  $\lambda = (\alpha^{2}/\mu \delta) g(\epsilon, dp/p_c)$
\\
  Price diffusion rate  & 
  $price^{2}/time$ & 
  $D \sim \mu^{2}\delta/\alpha$ & 
  ($\tau \rightarrow 0$, $dp \rightarrow 0$)~~$D_0 \sim 
    \mu^{2}\delta/\alpha\epsilon^{-0.5}$
\\ 
  & & & 
  ($\tau \rightarrow \infty$, $dp \rightarrow 0$) $D_\infty \sim \mu^{2}\delta
  /\alpha \epsilon^{0.5}$ 
\end{tabular}
\caption{Predictions of scaling of market properties vs. order flow.  The 
third column contains predictions from
the continuum analysis, in which the discreteness parameters are
ignored, and the fourth column gives more accurate predictions from theory
and simulation. The functions $f$ and $g$ are the order of magnitude of one throughout the
relevant ranges of variation of $\epsilon$ and $dp/p_c$.}
\label{scalingTable} 
\end{table*}

The bid-ask spread is the difference between the best price for buying
and selling.  It is an important determinant of transaction costs, as
it is the price that one pays to buy a share and then immediately sell
it.  The spread has dimensions of price and therefore scales under the
continuum approximation as $\mu / \alpha$.  Simulations and theory show
that the spread varies as $(\mu / \alpha) f(\epsilon, dp/p_c)$, where $f$
is a fairly flat function with $f(\epsilon, dp/p_c) \approx 1/2$
across much of the range of interest (see ref.~\cite{Smith02}).

\begin{figure}[ptb]
  \begin{center}
    \includegraphics[scale=0.37]{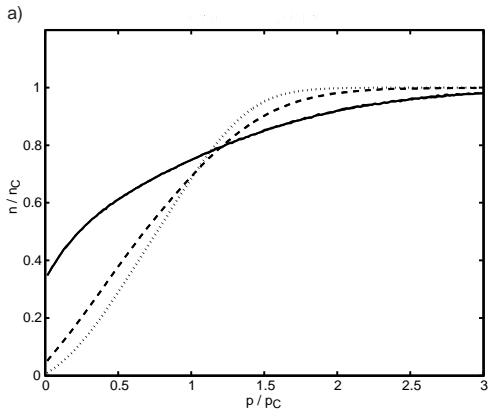}
    \caption{ 
      The mean depth profile versus price in nondimensional coordinates
      $\hat{n} = n p_c/N_c = n \delta / \alpha$ vs. 
      $\hat{p} = p/p_c = 2 \alpha p / \mu$.  The origin $p/p_c = 0$ 
      corresponds to the midpoint.  We show three different values of the 
      nondimensional granularity parameter: $\epsilon = 
      0.2$ (solid), $\epsilon = 0.02$ (dash), $\epsilon = 0.002$
      (dot), all with tick size $dp = 0$.  
    \label{epsDepth} 
    }
  \end{center}
\end{figure}

Another interesting quantity is the average {\it depth profile} $n(p)
= \langle n(p,t) \rangle$, which is the density of shares per price
interval.  The average depth profile is relatively small near the
midpoint and increases to an aysymptotic value far from the midpoint,
as shown in Fig.~\ref{epsDepth}.
The approach to an asymptotic value is a consequence of our assumption
of uniform order placement over an infinite range, and is not
realistic.  It should be viewed as a convenient boundary condition for
understanding the depth near $\hat{p} = 0$, where transactions occur.
From dimensional analysis the asymptotic depth, which has units of
{\it shares/price} is $\alpha/\delta$.  This result is exact.

An important property of the depth profile is its slope near the
origin, which determines the price response to the placement of a
small market order.  From continuum dimensional analysis the slope of the
average depth profile scales as $\lambda \sim \alpha^2 / \mu \delta$.
This is altered by effects due to the granularity of orders.  For
large $\epsilon$ the depth profile is a concave function with nonzero
values at $\hat{p} = 0$, whereas for small $\epsilon$, 
$n(0) \approx 0$ and $n(\hat{p})$ is convex near $\hat{p} = 0$.

In addition to the spread, the price response for executing a market
order is also a key factor determining transaction costs.  It can be
characterized by a price impact function $\Delta
p=\phi(\omega,\tau,t)$, where $\Delta p$ is the price shift at time
$t+\tau$ caused by a market order $\omega$ at time $t$.  Price impact
causes market friction, since selling tends to drive the price up and
buying tends to drive it down, so executing a circuit causes a loss.
(This loss, from changing the order distribution, is in addition to
the intrinsic loss created by a nonzero spread.)  The price impact is
closely related to the demand function, providing a natural starting
point for theories of statistical or dynamical properties of markets
\cite{Farmer98, Bouchaud98}. A naive argument predicts that the price impact 
$\phi (\omega)$ should increase at least linearly: Fractional price
changes should not depend on the scale of price. Suppose buying a
single share raises the price by a factor $k>1$. If $k$ is constant,
buying $\omega$ shares in succession should raise it by
$k^{\omega}$. Thus, if buying $\omega$ shares all at once affects the
price at least as much as buying them one at a time, the ratio of
prices before and after impact should increase at least exponentially.
Taking logarithms implies that the price impact as we have defined it
above should increase at least linearly.  In contrast, from empirical
studies $\phi(\omega)$ for buy orders grows more slowly than linearly
\cite{Hausman92,Farmer96,Torre97,Kempf98,Plerou01}.  
A recent more accurate study by Lillo et al. demonstrates that for the
New York Stock Exchange $\phi$ is strongly concave, and that in
general a simple power law of the form $\phi(\omega) \sim
\omega^{\beta}$ is not a good approximation \cite{Lillo02}.

\begin{figure}[ptb]
  \begin{center} 
    \includegraphics[scale=0.37]{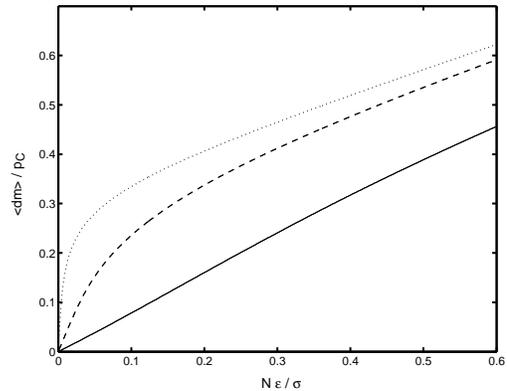}
  \caption{
    The average price impact corresponding to the results in
    Fig.~(\ref{epsDepth}). The average instantaneous movement of the
    nondimensional mid-price, $\langle dm \rangle/ p_c$ caused by an order 
    of size $N/N_c = N \epsilon/\sigma$.  $\epsilon = 0.2$ (solid),
    $\epsilon = 0.02$ (dash), $\epsilon = 0.002$ (dot).  
  \label{priceImpact}
  }
  \end{center}
\end{figure}

Our model gives average instantaneous price impact functions $\Delta
p=\langle \phi(\omega,0,t) \rangle$ as shown in Fig.~\ref{priceImpact}.
The price impact approaches a linear function for large $\epsilon$,
but for smaller values of $\epsilon$ it is strongly concave,
particularly near the midpoint.  Plotting this on log-log scale, this
function does not follow a pure power law.  For example, for $\epsilon
= 0.002$, the exponent is $\beta \approx 0.5$ for small orders, and
$\beta \approx 0.2$ for larger orders.  This is in agreement with the
results of Lillo et al. \cite{Lillo02}.  The instantaneous price
impact $\phi(\omega, 0, t)$ can be understood in terms of the depth
profile $n(p,t)$, as explained in ref.~\cite{Smith02}.

\begin{figure}[ptb] 
  \begin{center} %
  \includegraphics[scale=0.37]{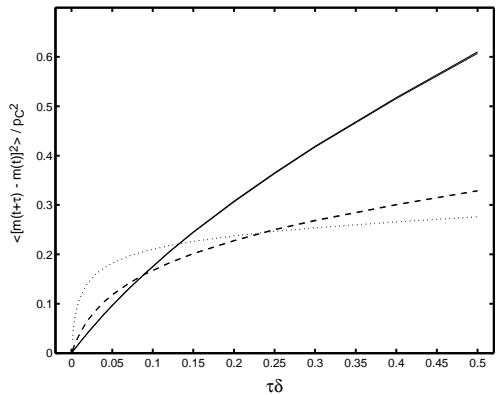}
    \caption{ 
      The variance of the change in the nondimensionalized midpoint
      price versus the nondimensional time lag.
      For a pure random walk this would be a straight line whose slope is 
      the diffusion rate, which is proportional to the square of the 
      volatility.  The 
      fact that the slope is steeper for short times comes from the nontrivial
      temporal persistence of orders.  The three cases
      correspond to Fig.~\ref{epsDepth}: $\epsilon = 0.2$ (solid),
      $\epsilon = 0.02$ (dash), $\epsilon = 0.002$ (dot).
    \label{epsVarVsTau}
    }
    \end{center}
\end{figure}

The price diffusion rate is a property of central interest.  In
finance, it is typically characterized in terms of the standard
deviation of prices at a particular timescale, which is referred to as
{\it volatility}.  Volatility is a measure of the uncertainty of price
movements and is the standard way to characterize risk.
In Fig.~\ref{epsVarVsTau} we plot simulation results for the
variance of the change in the midpoint price at timescale $\tau$, i.e.
the variance of $m(t+\tau) - m(t)$.
The slope is the diffusion rate, which at any fixed timescale is 
proportional to the square of the volatility. It appears that there
are at least two timescales involved, with a faster diffusion rate for
short timescales and a slower diffusion rate for long timescales. Such
correlated diffusion is not predicted by mean-field analysis.
Simulation results show that the diffusion rate is correctly described
by the product of the estimate from continuum dimensional analysis
$\mu^{2}\delta/{\alpha}^2$, and a
$\tau$-dependent power of the nondimensional granularity parameter
$\epsilon = 2 \delta\sigma/\mu$, as summarized in table
\ref{scalingTable}.  We cannot currently explain why this power
is $-1/2$ for short term diffusion and $1/2$ for long-term diffusion.

This model contains numerous simplifying assumptions.  Nevertheless,
it is the very simplicity of this model that allows us to make
unambiguous predictions about the most basic properties of real
markets.  Our prediction for the price impact function
agrees with the best empirical measurements and suggests
that concavity is a robust feature deriving from institutional
structure, rather than rationality
\cite{Lillo02}.  Preliminary analysis of data from the London Stock Exchange
using the nondimensional coordinates defined here shows an even better
collapse of the price impact function, suggesting the existence of
universal supply and demand functions.  Futhermore, the results
indicate that scaling relationships along the lines that we predict
have remarkable explanatory value for both the spread and the price
diffusion rate \cite{Daniels02}.  Even though we do not expect the
predictions of this model to be exact in every detail, they provide a
simple benchmark that can guide future improvements. Our model
illustrates how the need to store supply and demand gives rise to
interesting temporal properties of prices and liquidity even under
assumptions of perfectly random order flow, and demonstrates the
importance of making realistic models of market mechanisms.

\begin{acknowledgments}

We would like to thank the McKinsey Corporation, Credit Suisse First
Boston, Bob Maxfield, and Bill Miller for supporting this research.
G.~I.~thanks the Santa Fe Institute for kind hospitality.  We thank
Supriya Krishnamurthy for useful discussions.

\end{acknowledgments}

\end{document}